\documentstyle[12pt]{article}
\renewcommand{\baselinestretch}{1.5}
\begin{document}
\vskip 72pt
\centerline{\large TEMPERATURE DEPENDENCE OF HADRON MASSES FROM THE}
\centerline{\large ZIMANYI-MOSZKOWSKI MODEL}
\vskip 36pt
\centerline {\bf{Abhijit Bhattacharyya\footnote{Email : phys@boseinst.ernet.in}
and Sibaji Raha\footnote{Email : 
sibaji@boseinst.ernet.in}}}
\vskip 1pt
\centerline{Department of Physics, Bose Institute}
\centerline{93/1 A.P.C.Road}
\centerline{Calcutta 700 009, INDIA}
\vskip 36pt
\vskip 15 pt

\centerline{\bf {ABSTRACT}}
An extended version of the Zimanyi-Moskowski (ZM) model has been used to 
calculate the temperature dependence of the hadronic masses. The calculation 
of meson masses in the Random Phase Approximation (RPA) gives an increase in 
the $\sigma$, $\omega$ and $\pi$ effective masses with temperature. 


\newpage

In an earlier work \cite{a} we have calculated the density dependence of 
hadron masses from a newly proposed model: Zimanyi-Moskowski (ZM) model 
\cite{b}. The original ZM model was extended to include pions and the density 
dependence of the meson masses were calculated, from the extended model, in 
the Random Phase Approximation (RPA). This model differs from the popular 
Walecka model \cite{c} in the form of coupling between the scalar meson and 
the nucleon, which is a derivative coupling in the new model. As a result of 
this derivative coupling the model reproduces some of the experimental results 
nicely. It yields the incompressibility $K = 224.49 MeV$,  which  is 
    much closer to the experimental value $(K = 210 \pm 30 MeV)$ compared to 
    the Walecka model $(K = 524  MeV)$,  and  a  nucleon  effective  mass 
    $M_N^* = 797.64 MeV$ at the nuclear matter saturation density $(\rho = 0.17
fm^{-3})$. 
But, the price we have to pay is that the model looses its renormalisability 
due to the derivative coupling. This one can accept, at least as far as the 
discussion of the effective models goes, since the description of hadronic 
matter need only be valid upto the temperatures $T \le 200MeV$, provided the 
deconfinement phase transition to QCD matter is a reality.

    In \cite{a} we restricted ourselves to the zero temperature finite 
density scenario. But the temperature dependence of hadron masses, at zero 
density, is also very interesting to study as such a situation is expected to 
form in the central rapidity region of high energy heavy ion collisions. 
In this work, we would like to calculate the temperature dependence 
of hadron masses from the ZM model. We shall briefly describe the different 
variants  of the extended ZM model and then calculate the temperature 
dependence of hadron  masses.

The Lagrangians for the different variants of the ZM models along with the 
Walecka model can be written in a unified way, once we rescale different 
fields \cite{a,d}:
\newpage
\begin{eqnarray}
{\cal L}_R = {\bar \psi} i \gamma_\mu \partial^\mu \psi &-& {\bar \psi }(M_n -
m^{*\beta} g_\sigma \sigma) \psi - m^{* \alpha} \left[g_v{\bar \psi} 
\gamma_\mu \psi V^\mu - {1 \over 4}G_{\mu 
\nu} G^{\mu \nu} + {1 \over 2} m_v^2 V_\mu V^\mu \right] \nonumber\\
&+& {1 \over 2}(\partial_\mu \sigma \partial^\mu \sigma - m_\sigma^2 \sigma^2)
\end{eqnarray}

As in ref. \cite{a} we can extend the model to include pion and the total 
Lagrangian becomes 
\begin{equation}
{\cal L} = {\cal L}_R + {\cal L}_\pi
\end{equation}
where 
\begin{equation}
{\cal L}_\pi = {1 \over 2}(\partial_\mu \pi \partial^\mu \pi - m_\pi^2 \pi^2)
+ m^{*\beta} \left[{f_{\pi NN} \over m_\pi} \right] {\bar \psi} \gamma_\mu 
\gamma_5 \tau . \psi  \partial^\mu \pi
\end{equation}
where $\psi$, $\sigma$, $\pi$ and $V$  are respectively the nucleon, 
the scalar meson, the pion and the vector meson field; $M_N$, $m_\sigma$, 
$m_\pi$ and $m_v$ are the corresponding  masses; $g_\sigma$ and $g_v$ are the 
couplings of nucleon to scalar and vector mesons respectively; $G_{\mu \nu} = 
\partial_\mu V_\nu - \partial_\nu V_\mu$; $f_{\pi NN}$ the pion decay constant  
and $m^*$ is given as $m^* = (1+g_\sigma/M_N)^{-1}$. The parameters $\alpha$ 
and $\beta$ pick up the following values for different models: Walecka: 
$\alpha=0$, $\beta=0$; ZM: $\alpha=0$, $\beta=1$; ZM2: $\alpha=1$, $\beta=1$; 
ZM3: $\alpha=2$, $\beta=1$.

In the Mean Field Approximation (MFA), where the meson fields are replaced
by their ground state expectation values, the Lagrangian density is given  as 
\begin{equation}
{\cal L}_R^0 = {\bar \psi} \left[i \gamma . \partial - (M_N - m^{*\beta} 
g_\sigma \sigma_0) - m^{* \alpha} g_v \gamma_0  V^0 \right] \psi + m^{* \alpha} 
{1 \over 2} m_v^2 V_0^2 - {1 \over 2} m_\sigma^2 \sigma_0^2
\end{equation}

The energy density, calculated from ${\cal L}^0_R$, in the usual fashion, is 
given by 
\begin{equation}
{\cal E} = {g_v^2 \over {2 m_v^2}} m^{* \alpha} \rho^2_B + {m_\sigma^2 \over 
{2 g_\sigma^2}} \left[ {{1-m^*} \over m^{* \beta}} \right]^2 + {{2 \gamma} 
\over \pi^2} \int_0^{k_F} p^2 (p^2 + M_N^{*2})^{1/2} f(E_p^*) dp
\end{equation}

We fit the energy density to the nuclear ground state energy density at zero 
temperature  and nuclear saturation density to obtain the different coupling 
constants. The values of different parameters may be obtained in ref. 
\cite{a}. 

The nucleon effective mass at finite  temperature,  in  the  MFA, is given by 
\begin{equation}
M^*_N = M_N - g_\sigma m^{* 2\beta} \sigma_0  = M_N - {{4 M_N^* g^2_\sigma m^{* 2\beta}} 
\over {\pi^2 m_\sigma^2}} \int {{p^2 dp} \over {E_p^*}} f(E_p^* ) 
\end{equation}
    where $E^*_p  =\sqrt {p^2 + M_N^{*2}}$ is the effective energy  of  the  nucleon.  A 
    self consistent solution of the above equation gives the temperature 
    dependence of the effective nucleon mass. Such a calculation has also been
reported in ref. \cite{d} very recently. Now, using the ttemperature dependence 
of effective baryon mass we shall calculate the meson masses, at finite 
temperature, in the RPA.

    The  calculation  of  meson  masses  is  done  using  the  following 
    prescription : \cite{a}  the Dyson's equation relates the total Green's 
function $D(p)$ to the free Green's function $D_0(p)$ as
\begin{equation}
D^{-1}(p) = D^{-1}_0(p) + \Pi (p)
\end{equation}
    where $\Pi(p)$ is the polarisation function. The effect  of  interaction 
    is embedded in the polarisation  function.  The  pole  of  the  full 
    propagator then gives the effective mass. As in ref. \cite{a}, we restrict 
    ourselves to only the pole mass and ignore the screening  mass \cite{e}
    or the Landau mas. We have omitted the detailed calculation of the 
polarisation functions. The standard method for doing them can be obtained 
in ref. \cite{a}. In figures [1-3] we have plotted the temperature dependence 
of the effective meson masses.

The motivation behind this work was to look for the temperature dependence of 
the hadron masses from a class of newly proposed models, the Zimanyi-Moskowski 
models. In the ultrarelativistic heavy ion collisions, the 
matter formed in the central rapidity region is expected to be at a very high 
temperature and either at zero or very low baryon density. So the temperature 
dependence of hadron masses is important as far as the background of the QGP 
signal is concerned.
The ZM models differ from the usual Walecka model in the form of the coupling 
of the scalar mesons to the nucleons. The model has been extended to include 
the pseudoscalar pion which is the lightest of all mesons. Here we have 
incorporated, as in ref. \cite{a}, the pseudovector coupling of pions with 
nucleons. This model has been used to calculate the temperature dependence of 
the hadron masses. In the symmetric matter approximation, the equation of 
motion for the nucleon has been determined. This equation of motion has been 
used to calculate the nucleon propagator. We have used the nucleon propagator 
to calculate the nuclear energy density at the nuclear saturation density to 
calculate the coupling constants. The effective 
nucleon mass has been calculated in the MFA which is found to decrease with 
temperature for all the cases but, for the Walecka model the change is much 
more compared to the ZM models. We do not show this result here as they have 
also 
been reported by Delfino et.al. \cite{d} very recently. We, however, go much 
beyond this and use the  temperature  dependence  of 
    the nucleon mass to calculate  the  temperature  dependence  of  the 
    meson masses, in the RPA, self consistently. The $\sigma$, $\omega$ and 
$\pi$ masses 
    are found to increase with temperature, both in the ZM and Walecka models.
Special attention may be paid  to  the  pion 
    mass. The main problem of treating pions in  the  Walecka  model  is 
    that its mass increases with temperature abnormally as seen from fig.~3. 
One, however, expects that the pion mass should not  change  much,  as  the 
    pion, being a Goldstone boson, is protected by chiral symmetry \cite{g}. 
    As seen from fig.3 here, the pion mass in the  ZM  models  does  not 
    change much with temperature and one can expect that the problem  of 
    abnormal change in pion mass,  which  is  observed  in  the  Walecka 
    model, can be overcome in the ZM models.  It  should  of  course  be 
    mentioned in this context that neither the Walecka model nor the  ZM 
    models incorporate chiral symmetry and thus the  preceding  argument 
    is somewhat extraneous. The $\sigma$-meson mass should decrease in a 
chiral model and becomes degenerate with the pion mass as a signature of 
the restoration of chiral symmetry at high temperature.
But, as far as only the behaviour of the pion is concerned, a model of nuclear 
matter with 
    "correct" behaviour of the pion mass is quite appealing and in  this 
    respect the ZM models appear to perform better. If one  carries  the 
    calculation beyond the temperature range of $200MeV$ the same  problem 
    might appear in the ZM models also. But, at  that  high  temperature 
    the deconfinement phase transition is expected to occur and  hadrons 
    are no longer the elementary particles. Hence the results at such  a 
    high temperature range are not reliable at all.

    So, to conclude, the temperature  dependence  of  hadron  masses  is 
    still a very controversial subject \cite{h}.  In 
    this specific model, the Zimanyi-Moskowski  model,  the  temperature 
    dependence of the meson masses, especially the pion, comes out to be 
    much more realistic compared to other models. The present model  can 
    be  extended  to  include  $\rho$  and  $a_1$   mesons  and  the  
meson-meson couplings. Such calculations are in  progress  and 
    will be comunicated shortly.

The work of Abhijit Bhattacharyya has been supported partially by 
    the Department of Atomic Energy (Government of  India).

\newpage
\begin{figure}
% GNUPLOT: LaTeX picture
\setlength{\unitlength}{0.240900pt}
\ifx\plotpoint\undefined\newsavebox{\plotpoint}\fi
\sbox{\plotpoint}{\rule[-0.500pt]{1.000pt}{1.000pt}}%
\begin{picture}(1500,1049)(0,0)
\font\gnuplot=cmr10 at 10pt
\gnuplot
\sbox{\plotpoint}{\rule[-0.500pt]{1.000pt}{1.000pt}}%
\put(176.0,68.0){\rule[-0.500pt]{1.000pt}{230.782pt}}
\put(176.0,68.0){\rule[-0.500pt]{4.818pt}{1.000pt}}
\put(154,68){\makebox(0,0)[r]{$520$}}
\put(1416.0,68.0){\rule[-0.500pt]{4.818pt}{1.000pt}}
\put(176.0,215.0){\rule[-0.500pt]{4.818pt}{1.000pt}}
\put(154,215){\makebox(0,0)[r]{$540$}}
\put(1416.0,215.0){\rule[-0.500pt]{4.818pt}{1.000pt}}
\put(176.0,362.0){\rule[-0.500pt]{4.818pt}{1.000pt}}
\put(154,362){\makebox(0,0)[r]{$560$}}
\put(1416.0,362.0){\rule[-0.500pt]{4.818pt}{1.000pt}}
\put(176.0,509.0){\rule[-0.500pt]{4.818pt}{1.000pt}}
\put(154,509){\makebox(0,0)[r]{$580$}}
\put(1416.0,509.0){\rule[-0.500pt]{4.818pt}{1.000pt}}
\put(176.0,656.0){\rule[-0.500pt]{4.818pt}{1.000pt}}
\put(154,656){\makebox(0,0)[r]{$600$}}
\put(1416.0,656.0){\rule[-0.500pt]{4.818pt}{1.000pt}}
\put(176.0,803.0){\rule[-0.500pt]{4.818pt}{1.000pt}}
\put(154,803){\makebox(0,0)[r]{$620$}}
\put(1416.0,803.0){\rule[-0.500pt]{4.818pt}{1.000pt}}
\put(176.0,950.0){\rule[-0.500pt]{4.818pt}{1.000pt}}
\put(154,950){\makebox(0,0)[r]{$640$}}
\put(1416.0,950.0){\rule[-0.500pt]{4.818pt}{1.000pt}}
\put(176.0,68.0){\rule[-0.500pt]{1.000pt}{4.818pt}}
\put(176,23){\makebox(0,0){$0$}}
\put(176.0,1006.0){\rule[-0.500pt]{1.000pt}{4.818pt}}
\put(490.0,68.0){\rule[-0.500pt]{1.000pt}{4.818pt}}
\put(490,23){\makebox(0,0){$50$}}
\put(490.0,1006.0){\rule[-0.500pt]{1.000pt}{4.818pt}}
\put(804.0,68.0){\rule[-0.500pt]{1.000pt}{4.818pt}}
\put(804,23){\makebox(0,0){$100$}}
\put(804.0,1006.0){\rule[-0.500pt]{1.000pt}{4.818pt}}
\put(1119.0,68.0){\rule[-0.500pt]{1.000pt}{4.818pt}}
\put(1119,23){\makebox(0,0){$150$}}
\put(1119.0,1006.0){\rule[-0.500pt]{1.000pt}{4.818pt}}
\put(1433.0,68.0){\rule[-0.500pt]{1.000pt}{4.818pt}}
\put(1433,23){\makebox(0,0){$200$}}
\put(1433.0,1006.0){\rule[-0.500pt]{1.000pt}{4.818pt}}
\put(176.0,68.0){\rule[-0.500pt]{303.534pt}{1.000pt}}
\put(1436.0,68.0){\rule[-0.500pt]{1.000pt}{230.782pt}}
\put(176.0,1026.0){\rule[-0.500pt]{303.534pt}{1.000pt}}
\put(911,-78){\makebox(0,0)[r]{$T(MeV)$}}
\put(1875,-225){\makebox(0,0)[r]{{\bf{Figure 1: Temperature dependence of $\sigma$-meson mass; W refers to the Walecka model}}}}
\put(1950,-300){\makebox(0,0)[r]{{\bf{and ZM, ZM2 and ZM3 refer to the different versions of the Zimanyi-Moskowski model.}}}}
\put(50,509){\makebox(0,0)[r]{$m_\sigma^*(MeV)$}}
\put(1313,803){\makebox(0,0)[r]{$W$}}
\put(1401,325){\makebox(0,0)[r]{$ZM$}}
\put(1401,435){\makebox(0,0)[r]{$ZM2$}}
\put(1433,619){\makebox(0,0)[r]{$ZM3$}}
\put(176.0,68.0){\rule[-0.500pt]{1.000pt}{230.782pt}}
\sbox{\plotpoint}{\rule[-0.300pt]{0.600pt}{0.600pt}}%
\put(176,288){\usebox{\plotpoint}}
\put(773,287.25){\rule{7.468pt}{0.600pt}}
\multiput(773.00,286.75)(15.500,1.000){2}{\rule{3.734pt}{0.600pt}}
\put(176.0,288.0){\rule[-0.300pt]{143.817pt}{0.600pt}}
\put(836,288.25){\rule{7.468pt}{0.600pt}}
\multiput(836.00,287.75)(15.500,1.000){2}{\rule{3.734pt}{0.600pt}}
\put(867,289.25){\rule{7.709pt}{0.600pt}}
\multiput(867.00,288.75)(16.000,1.000){2}{\rule{3.854pt}{0.600pt}}
\put(899,290.75){\rule{7.468pt}{0.600pt}}
\multiput(899.00,289.75)(15.500,2.000){2}{\rule{3.734pt}{0.600pt}}
\put(930,292.75){\rule{7.709pt}{0.600pt}}
\multiput(930.00,291.75)(16.000,2.000){2}{\rule{3.854pt}{0.600pt}}
\put(962,295.25){\rule{6.350pt}{0.600pt}}
\multiput(962.00,293.75)(17.820,3.000){2}{\rule{3.175pt}{0.600pt}}
\multiput(993.00,298.99)(5.537,0.503){3}{\rule{4.800pt}{0.121pt}}
\multiput(993.00,296.75)(21.037,4.000){2}{\rule{2.400pt}{0.600pt}}
\multiput(1024.00,302.99)(3.859,0.502){5}{\rule{3.990pt}{0.121pt}}
\multiput(1024.00,300.75)(23.719,5.000){2}{\rule{1.995pt}{0.600pt}}
\multiput(1056.00,307.99)(1.817,0.501){13}{\rule{2.217pt}{0.121pt}}
\multiput(1056.00,305.75)(26.399,9.000){2}{\rule{1.108pt}{0.600pt}}
\multiput(1087.00,317.00)(1.672,0.501){15}{\rule{2.070pt}{0.121pt}}
\multiput(1087.00,314.75)(27.704,10.000){2}{\rule{1.035pt}{0.600pt}}
\multiput(1119.00,327.00)(1.130,0.500){23}{\rule{1.479pt}{0.121pt}}
\multiput(1119.00,324.75)(27.931,14.000){2}{\rule{0.739pt}{0.600pt}}
\multiput(1150.00,341.00)(0.780,0.500){35}{\rule{1.080pt}{0.121pt}}
\multiput(1150.00,338.75)(28.758,20.000){2}{\rule{0.540pt}{0.600pt}}
\multiput(1181.00,361.00)(0.592,0.500){49}{\rule{0.861pt}{0.120pt}}
\multiput(1181.00,358.75)(30.213,27.000){2}{\rule{0.431pt}{0.600pt}}
\multiput(1214.00,387.00)(0.500,0.629){57}{\rule{0.120pt}{0.905pt}}
\multiput(1211.75,387.00)(31.000,37.122){2}{\rule{0.600pt}{0.452pt}}
\multiput(1245.00,426.00)(0.500,0.943){59}{\rule{0.120pt}{1.275pt}}
\multiput(1242.75,426.00)(32.000,57.354){2}{\rule{0.600pt}{0.638pt}}
\multiput(1277.00,486.00)(0.500,1.713){57}{\rule{0.120pt}{2.182pt}}
\multiput(1274.75,486.00)(31.000,100.471){2}{\rule{0.600pt}{1.091pt}}
\multiput(1308.00,591.00)(0.500,4.155){59}{\rule{0.120pt}{5.063pt}}
\multiput(1305.75,591.00)(32.000,251.493){2}{\rule{0.600pt}{2.531pt}}
\put(804.0,289.0){\rule[-0.300pt]{7.709pt}{0.600pt}}
\put(176,288){\usebox{\plotpoint}}
\put(804,287.25){\rule{7.709pt}{0.600pt}}
\multiput(804.00,286.75)(16.000,1.000){2}{\rule{3.854pt}{0.600pt}}
\put(176.0,288.0){\rule[-0.300pt]{151.285pt}{0.600pt}}
\put(867,288.25){\rule{7.709pt}{0.600pt}}
\multiput(867.00,287.75)(16.000,1.000){2}{\rule{3.854pt}{0.600pt}}
\put(899,289.25){\rule{7.468pt}{0.600pt}}
\multiput(899.00,288.75)(15.500,1.000){2}{\rule{3.734pt}{0.600pt}}
\put(836.0,289.0){\rule[-0.300pt]{7.468pt}{0.600pt}}
\put(962,290.75){\rule{7.468pt}{0.600pt}}
\multiput(962.00,289.75)(15.500,2.000){2}{\rule{3.734pt}{0.600pt}}
\put(993,292.25){\rule{7.468pt}{0.600pt}}
\multiput(993.00,291.75)(15.500,1.000){2}{\rule{3.734pt}{0.600pt}}
\put(1024,294.25){\rule{6.550pt}{0.600pt}}
\multiput(1024.00,292.75)(18.405,3.000){2}{\rule{3.275pt}{0.600pt}}
\put(1056,297.25){\rule{6.350pt}{0.600pt}}
\multiput(1056.00,295.75)(17.820,3.000){2}{\rule{3.175pt}{0.600pt}}
\multiput(1087.00,300.99)(5.726,0.503){3}{\rule{4.950pt}{0.121pt}}
\multiput(1087.00,298.75)(21.726,4.000){2}{\rule{2.475pt}{0.600pt}}
\multiput(1119.00,304.99)(3.734,0.502){5}{\rule{3.870pt}{0.121pt}}
\multiput(1119.00,302.75)(22.968,5.000){2}{\rule{1.935pt}{0.600pt}}
\multiput(1150.00,309.99)(2.418,0.501){9}{\rule{2.807pt}{0.121pt}}
\multiput(1150.00,307.75)(25.174,7.000){2}{\rule{1.404pt}{0.600pt}}
\multiput(1181.00,316.99)(2.141,0.501){11}{\rule{2.550pt}{0.121pt}}
\multiput(1181.00,314.75)(26.707,8.000){2}{\rule{1.275pt}{0.600pt}}
\multiput(1213.00,325.00)(1.618,0.501){15}{\rule{2.010pt}{0.121pt}}
\multiput(1213.00,322.75)(26.828,10.000){2}{\rule{1.005pt}{0.600pt}}
\multiput(1244.00,335.00)(1.374,0.500){19}{\rule{1.750pt}{0.121pt}}
\multiput(1244.00,332.75)(28.368,12.000){2}{\rule{0.875pt}{0.600pt}}
\multiput(1276.00,347.00)(1.130,0.500){23}{\rule{1.479pt}{0.121pt}}
\multiput(1276.00,344.75)(27.931,14.000){2}{\rule{0.739pt}{0.600pt}}
\multiput(1307.00,361.00)(0.898,0.500){31}{\rule{1.217pt}{0.121pt}}
\multiput(1307.00,358.75)(29.475,18.000){2}{\rule{0.608pt}{0.600pt}}
\put(930.0,291.0){\rule[-0.300pt]{7.709pt}{0.600pt}}
\put(176,288){\usebox{\plotpoint}}
\put(804,287.25){\rule{7.709pt}{0.600pt}}
\multiput(804.00,286.75)(16.000,1.000){2}{\rule{3.854pt}{0.600pt}}
\put(176.0,288.0){\rule[-0.300pt]{151.285pt}{0.600pt}}
\put(867,288.25){\rule{7.709pt}{0.600pt}}
\multiput(867.00,287.75)(16.000,1.000){2}{\rule{3.854pt}{0.600pt}}
\put(899,289.25){\rule{7.468pt}{0.600pt}}
\multiput(899.00,288.75)(15.500,1.000){2}{\rule{3.734pt}{0.600pt}}
\put(930,290.25){\rule{7.709pt}{0.600pt}}
\multiput(930.00,289.75)(16.000,1.000){2}{\rule{3.854pt}{0.600pt}}
\put(962,291.75){\rule{7.468pt}{0.600pt}}
\multiput(962.00,290.75)(15.500,2.000){2}{\rule{3.734pt}{0.600pt}}
\put(993,293.75){\rule{7.468pt}{0.600pt}}
\multiput(993.00,292.75)(15.500,2.000){2}{\rule{3.734pt}{0.600pt}}
\put(1024,296.25){\rule{6.550pt}{0.600pt}}
\multiput(1024.00,294.75)(18.405,3.000){2}{\rule{3.275pt}{0.600pt}}
\multiput(1056.00,299.99)(3.734,0.502){5}{\rule{3.870pt}{0.121pt}}
\multiput(1056.00,297.75)(22.968,5.000){2}{\rule{1.935pt}{0.600pt}}
\multiput(1087.00,304.99)(3.859,0.502){5}{\rule{3.990pt}{0.121pt}}
\multiput(1087.00,302.75)(23.719,5.000){2}{\rule{1.995pt}{0.600pt}}
\multiput(1119.00,309.99)(2.418,0.501){9}{\rule{2.807pt}{0.121pt}}
\multiput(1119.00,307.75)(25.174,7.000){2}{\rule{1.404pt}{0.600pt}}
\multiput(1150.00,316.99)(2.073,0.501){11}{\rule{2.475pt}{0.121pt}}
\multiput(1150.00,314.75)(25.863,8.000){2}{\rule{1.238pt}{0.600pt}}
\multiput(1181.00,325.00)(1.508,0.501){17}{\rule{1.895pt}{0.121pt}}
\multiput(1181.00,322.75)(28.066,11.000){2}{\rule{0.948pt}{0.600pt}}
\multiput(1213.00,336.00)(1.130,0.500){23}{\rule{1.479pt}{0.121pt}}
\multiput(1213.00,333.75)(27.931,14.000){2}{\rule{0.739pt}{0.600pt}}
\multiput(1244.00,350.00)(0.953,0.500){29}{\rule{1.279pt}{0.121pt}}
\multiput(1244.00,347.75)(29.345,17.000){2}{\rule{0.640pt}{0.600pt}}
\multiput(1276.00,367.00)(0.780,0.500){35}{\rule{1.080pt}{0.121pt}}
\multiput(1276.00,364.75)(28.758,20.000){2}{\rule{0.540pt}{0.600pt}}
\multiput(1307.00,387.00)(0.668,0.500){43}{\rule{0.950pt}{0.121pt}}
\multiput(1307.00,384.75)(30.028,24.000){2}{\rule{0.475pt}{0.600pt}}
\put(836.0,289.0){\rule[-0.300pt]{7.468pt}{0.600pt}}
\put(176,288){\usebox{\plotpoint}}
\put(742,287.25){\rule{7.468pt}{0.600pt}}
\multiput(742.00,286.75)(15.500,1.000){2}{\rule{3.734pt}{0.600pt}}
\put(176.0,288.0){\rule[-0.300pt]{136.349pt}{0.600pt}}
\put(804,288.25){\rule{7.709pt}{0.600pt}}
\multiput(804.00,287.75)(16.000,1.000){2}{\rule{3.854pt}{0.600pt}}
\put(836,289.25){\rule{7.468pt}{0.600pt}}
\multiput(836.00,288.75)(15.500,1.000){2}{\rule{3.734pt}{0.600pt}}
\put(773.0,289.0){\rule[-0.300pt]{7.468pt}{0.600pt}}
\put(899,291.25){\rule{6.350pt}{0.600pt}}
\multiput(899.00,289.75)(17.820,3.000){2}{\rule{3.175pt}{0.600pt}}
\put(930,293.75){\rule{7.709pt}{0.600pt}}
\multiput(930.00,292.75)(16.000,2.000){2}{\rule{3.854pt}{0.600pt}}
\multiput(962.00,296.99)(5.537,0.503){3}{\rule{4.800pt}{0.121pt}}
\multiput(962.00,294.75)(21.037,4.000){2}{\rule{2.400pt}{0.600pt}}
\multiput(993.00,300.99)(3.734,0.502){5}{\rule{3.870pt}{0.121pt}}
\multiput(993.00,298.75)(22.968,5.000){2}{\rule{1.935pt}{0.600pt}}
\multiput(1024.00,305.99)(2.499,0.501){9}{\rule{2.893pt}{0.121pt}}
\multiput(1024.00,303.75)(25.996,7.000){2}{\rule{1.446pt}{0.600pt}}
\multiput(1056.00,312.99)(1.817,0.501){13}{\rule{2.217pt}{0.121pt}}
\multiput(1056.00,310.75)(26.399,9.000){2}{\rule{1.108pt}{0.600pt}}
\multiput(1087.00,322.00)(1.374,0.500){19}{\rule{1.750pt}{0.121pt}}
\multiput(1087.00,319.75)(28.368,12.000){2}{\rule{0.875pt}{0.600pt}}
\multiput(1119.00,334.00)(0.983,0.500){27}{\rule{1.313pt}{0.121pt}}
\multiput(1119.00,331.75)(28.276,16.000){2}{\rule{0.656pt}{0.600pt}}
\multiput(1150.00,350.00)(0.742,0.500){37}{\rule{1.036pt}{0.121pt}}
\multiput(1150.00,347.75)(28.850,21.000){2}{\rule{0.518pt}{0.600pt}}
\multiput(1181.00,371.00)(0.615,0.500){47}{\rule{0.888pt}{0.120pt}}
\multiput(1181.00,368.75)(30.156,26.000){2}{\rule{0.444pt}{0.600pt}}
\multiput(1214.00,396.00)(0.500,0.547){57}{\rule{0.120pt}{0.808pt}}
\multiput(1211.75,396.00)(31.000,32.323){2}{\rule{0.600pt}{0.404pt}}
\multiput(1245.00,430.00)(0.500,0.673){59}{\rule{0.120pt}{0.956pt}}
\multiput(1242.75,430.00)(32.000,41.015){2}{\rule{0.600pt}{0.478pt}}
\multiput(1277.00,473.00)(0.500,0.882){69}{\rule{0.120pt}{1.204pt}}
\multiput(1274.75,473.00)(37.000,62.501){2}{\rule{0.600pt}{0.602pt}}
\multiput(1314.00,538.00)(0.500,1.028){47}{\rule{0.120pt}{1.373pt}}
\multiput(1311.75,538.00)(26.000,50.150){2}{\rule{0.600pt}{0.687pt}}
\put(867.0,291.0){\rule[-0.300pt]{7.709pt}{0.600pt}}
\end{picture}
\end{figure}
\newpage
\begin{figure}
% GNUPLOT: LaTeX picture
\setlength{\unitlength}{0.240900pt}
\ifx\plotpoint\undefined\newsavebox{\plotpoint}\fi
\sbox{\plotpoint}{\rule[-0.500pt]{1.000pt}{1.000pt}}%
\begin{picture}(1500,1049)(0,0)
\font\gnuplot=cmr10 at 10pt
\gnuplot
\sbox{\plotpoint}{\rule[-0.500pt]{1.000pt}{1.000pt}}%
\put(176.0,68.0){\rule[-0.500pt]{1.000pt}{230.782pt}}
\put(176.0,68.0){\rule[-0.500pt]{4.818pt}{1.000pt}}
\put(154,68){\makebox(0,0)[r]{$750$}}
\put(1416.0,68.0){\rule[-0.500pt]{4.818pt}{1.000pt}}
\put(176.0,227.0){\rule[-0.500pt]{4.818pt}{1.000pt}}
\put(154,227){\makebox(0,0)[r]{$800$}}
\put(1416.0,227.0){\rule[-0.500pt]{4.818pt}{1.000pt}}
\put(176.0,387.0){\rule[-0.500pt]{4.818pt}{1.000pt}}
\put(154,387){\makebox(0,0)[r]{$850$}}
\put(1416.0,387.0){\rule[-0.500pt]{4.818pt}{1.000pt}}
\put(176.0,546.0){\rule[-0.500pt]{4.818pt}{1.000pt}}
\put(154,546){\makebox(0,0)[r]{$900$}}
\put(1416.0,546.0){\rule[-0.500pt]{4.818pt}{1.000pt}}
\put(176.0,706.0){\rule[-0.500pt]{4.818pt}{1.000pt}}
\put(154,706){\makebox(0,0)[r]{$950$}}
\put(1416.0,706.0){\rule[-0.500pt]{4.818pt}{1.000pt}}
\put(176.0,865.0){\rule[-0.500pt]{4.818pt}{1.000pt}}
\put(154,865){\makebox(0,0)[r]{$1000$}}
\put(1416.0,865.0){\rule[-0.500pt]{4.818pt}{1.000pt}}
\put(176.0,1025.0){\rule[-0.500pt]{4.818pt}{1.000pt}}
\put(154,1025){\makebox(0,0)[r]{$1050$}}
\put(1416.0,1025.0){\rule[-0.500pt]{4.818pt}{1.000pt}}
\put(176.0,68.0){\rule[-0.500pt]{1.000pt}{4.818pt}}
\put(176,23){\makebox(0,0){$0$}}
\put(176.0,1006.0){\rule[-0.500pt]{1.000pt}{4.818pt}}
\put(490.0,68.0){\rule[-0.500pt]{1.000pt}{4.818pt}}
\put(490,23){\makebox(0,0){$50$}}
\put(490.0,1006.0){\rule[-0.500pt]{1.000pt}{4.818pt}}
\put(804.0,68.0){\rule[-0.500pt]{1.000pt}{4.818pt}}
\put(804,23){\makebox(0,0){$100$}}
\put(804.0,1006.0){\rule[-0.500pt]{1.000pt}{4.818pt}}
\put(1119.0,68.0){\rule[-0.500pt]{1.000pt}{4.818pt}}
\put(1119,23){\makebox(0,0){$150$}}
\put(1119.0,1006.0){\rule[-0.500pt]{1.000pt}{4.818pt}}
\put(1433.0,68.0){\rule[-0.500pt]{1.000pt}{4.818pt}}
\put(1433,23){\makebox(0,0){$200$}}
\put(1433.0,1006.0){\rule[-0.500pt]{1.000pt}{4.818pt}}
\put(176.0,68.0){\rule[-0.500pt]{303.534pt}{1.000pt}}
\put(1436.0,68.0){\rule[-0.500pt]{1.000pt}{230.782pt}}
\put(176.0,1026.0){\rule[-0.500pt]{303.534pt}{1.000pt}}
\put(911,-43){\makebox(0,0)[r]{$T(MeV)$}}
\put(1950,-170){\makebox(0,0)[r]{{\bf{Figure 2: Temperature dependence of $\omega$-meson mass; the nomenclature is same as fig.1.}}}}
\put(50,546){\makebox(0,0)[r]{$m_\omega^*(MeV)$}}
\put(1313,897){\makebox(0,0)[r]{$W$}}
\put(1401,160){\makebox(0,0)[r]{$ZM$}}
\put(1401,275){\makebox(0,0)[r]{$ZM2$}}
\put(1433,403){\makebox(0,0)[r]{$ZM3$}}
\put(176.0,68.0){\rule[-0.500pt]{1.000pt}{230.782pt}}
\sbox{\plotpoint}{\rule[-0.300pt]{0.600pt}{0.600pt}}%
\put(176,173){\usebox{\plotpoint}}
\put(679,172.25){\rule{7.468pt}{0.600pt}}
\multiput(679.00,171.75)(15.500,1.000){2}{\rule{3.734pt}{0.600pt}}
\put(176.0,173.0){\rule[-0.300pt]{121.173pt}{0.600pt}}
\put(804,173.25){\rule{7.709pt}{0.600pt}}
\multiput(804.00,172.75)(16.000,1.000){2}{\rule{3.854pt}{0.600pt}}
\put(836,174.25){\rule{7.468pt}{0.600pt}}
\multiput(836.00,173.75)(15.500,1.000){2}{\rule{3.734pt}{0.600pt}}
\put(867,175.75){\rule{7.709pt}{0.600pt}}
\multiput(867.00,174.75)(16.000,2.000){2}{\rule{3.854pt}{0.600pt}}
\put(899,177.75){\rule{7.468pt}{0.600pt}}
\multiput(899.00,176.75)(15.500,2.000){2}{\rule{3.734pt}{0.600pt}}
\put(930,180.25){\rule{6.550pt}{0.600pt}}
\multiput(930.00,178.75)(18.405,3.000){2}{\rule{3.275pt}{0.600pt}}
\multiput(962.00,183.99)(5.537,0.503){3}{\rule{4.800pt}{0.121pt}}
\multiput(962.00,181.75)(21.037,4.000){2}{\rule{2.400pt}{0.600pt}}
\multiput(993.00,187.99)(3.734,0.502){5}{\rule{3.870pt}{0.121pt}}
\multiput(993.00,185.75)(22.968,5.000){2}{\rule{1.935pt}{0.600pt}}
\multiput(1024.00,192.99)(2.141,0.501){11}{\rule{2.550pt}{0.121pt}}
\multiput(1024.00,190.75)(26.707,8.000){2}{\rule{1.275pt}{0.600pt}}
\multiput(1056.00,200.99)(1.817,0.501){13}{\rule{2.217pt}{0.121pt}}
\multiput(1056.00,198.75)(26.399,9.000){2}{\rule{1.108pt}{0.600pt}}
\multiput(1087.00,210.00)(1.374,0.500){19}{\rule{1.750pt}{0.121pt}}
\multiput(1087.00,207.75)(28.368,12.000){2}{\rule{0.875pt}{0.600pt}}
\multiput(1119.00,222.00)(0.983,0.500){27}{\rule{1.313pt}{0.121pt}}
\multiput(1119.00,219.75)(28.276,16.000){2}{\rule{0.656pt}{0.600pt}}
\multiput(1150.00,238.00)(0.780,0.500){35}{\rule{1.080pt}{0.121pt}}
\multiput(1150.00,235.75)(28.758,20.000){2}{\rule{0.540pt}{0.600pt}}
\multiput(1181.00,258.00)(0.571,0.500){51}{\rule{0.836pt}{0.120pt}}
\multiput(1181.00,255.75)(30.265,28.000){2}{\rule{0.418pt}{0.600pt}}
\multiput(1214.00,285.00)(0.500,0.629){57}{\rule{0.120pt}{0.905pt}}
\multiput(1211.75,285.00)(31.000,37.122){2}{\rule{0.600pt}{0.452pt}}
\multiput(1245.00,324.00)(0.500,0.927){59}{\rule{0.120pt}{1.256pt}}
\multiput(1242.75,324.00)(32.000,56.393){2}{\rule{0.600pt}{0.628pt}}
\multiput(1277.00,383.00)(0.500,1.648){57}{\rule{0.120pt}{2.105pt}}
\multiput(1274.75,383.00)(31.000,96.631){2}{\rule{0.600pt}{1.052pt}}
\multiput(1308.00,484.00)(0.500,4.250){59}{\rule{0.120pt}{5.175pt}}
\multiput(1305.75,484.00)(32.000,257.259){2}{\rule{0.600pt}{2.588pt}}
\multiput(1339.99,752.00)(0.501,17.638){7}{\rule{0.121pt}{18.350pt}}
\multiput(1337.75,752.00)(6.000,143.914){2}{\rule{0.600pt}{9.175pt}}
\put(710.0,174.0){\rule[-0.300pt]{22.645pt}{0.600pt}}
\put(176,173){\usebox{\plotpoint}}
\put(773,172.25){\rule{7.468pt}{0.600pt}}
\multiput(773.00,171.75)(15.500,1.000){2}{\rule{3.734pt}{0.600pt}}
\put(176.0,173.0){\rule[-0.300pt]{143.817pt}{0.600pt}}
\put(899,173.25){\rule{7.468pt}{0.600pt}}
\multiput(899.00,172.75)(15.500,1.000){2}{\rule{3.734pt}{0.600pt}}
\put(804.0,174.0){\rule[-0.300pt]{22.885pt}{0.600pt}}
\put(962,174.25){\rule{7.468pt}{0.600pt}}
\multiput(962.00,173.75)(15.500,1.000){2}{\rule{3.734pt}{0.600pt}}
\put(993,175.25){\rule{7.468pt}{0.600pt}}
\multiput(993.00,174.75)(15.500,1.000){2}{\rule{3.734pt}{0.600pt}}
\put(1024,176.75){\rule{7.709pt}{0.600pt}}
\multiput(1024.00,175.75)(16.000,2.000){2}{\rule{3.854pt}{0.600pt}}
\put(1056,178.25){\rule{7.468pt}{0.600pt}}
\multiput(1056.00,177.75)(15.500,1.000){2}{\rule{3.734pt}{0.600pt}}
\put(1087,179.75){\rule{7.709pt}{0.600pt}}
\multiput(1087.00,178.75)(16.000,2.000){2}{\rule{3.854pt}{0.600pt}}
\put(1119,182.25){\rule{6.350pt}{0.600pt}}
\multiput(1119.00,180.75)(17.820,3.000){2}{\rule{3.175pt}{0.600pt}}
\multiput(1150.00,185.99)(5.537,0.503){3}{\rule{4.800pt}{0.121pt}}
\multiput(1150.00,183.75)(21.037,4.000){2}{\rule{2.400pt}{0.600pt}}
\multiput(1181.00,189.99)(5.726,0.503){3}{\rule{4.950pt}{0.121pt}}
\multiput(1181.00,187.75)(21.726,4.000){2}{\rule{2.475pt}{0.600pt}}
\multiput(1213.00,193.99)(3.734,0.502){5}{\rule{3.870pt}{0.121pt}}
\multiput(1213.00,191.75)(22.968,5.000){2}{\rule{1.935pt}{0.600pt}}
\multiput(1244.00,198.99)(3.016,0.501){7}{\rule{3.350pt}{0.121pt}}
\multiput(1244.00,196.75)(25.047,6.000){2}{\rule{1.675pt}{0.600pt}}
\multiput(1276.00,204.99)(2.418,0.501){9}{\rule{2.807pt}{0.121pt}}
\multiput(1276.00,202.75)(25.174,7.000){2}{\rule{1.404pt}{0.600pt}}
\multiput(1307.00,211.99)(2.141,0.501){11}{\rule{2.550pt}{0.121pt}}
\multiput(1307.00,209.75)(26.707,8.000){2}{\rule{1.275pt}{0.600pt}}
\multiput(1339.00,220.00)(1.618,0.501){15}{\rule{2.010pt}{0.121pt}}
\multiput(1339.00,217.75)(26.828,10.000){2}{\rule{1.005pt}{0.600pt}}
\put(930.0,175.0){\rule[-0.300pt]{7.709pt}{0.600pt}}
\put(176,173){\usebox{\plotpoint}}
\put(742,172.25){\rule{7.468pt}{0.600pt}}
\multiput(742.00,171.75)(15.500,1.000){2}{\rule{3.734pt}{0.600pt}}
\put(176.0,173.0){\rule[-0.300pt]{136.349pt}{0.600pt}}
\put(867,173.25){\rule{7.709pt}{0.600pt}}
\multiput(867.00,172.75)(16.000,1.000){2}{\rule{3.854pt}{0.600pt}}
\put(899,174.25){\rule{7.468pt}{0.600pt}}
\multiput(899.00,173.75)(15.500,1.000){2}{\rule{3.734pt}{0.600pt}}
\put(930,175.25){\rule{7.709pt}{0.600pt}}
\multiput(930.00,174.75)(16.000,1.000){2}{\rule{3.854pt}{0.600pt}}
\put(962,176.25){\rule{7.468pt}{0.600pt}}
\multiput(962.00,175.75)(15.500,1.000){2}{\rule{3.734pt}{0.600pt}}
\put(993,177.75){\rule{7.468pt}{0.600pt}}
\multiput(993.00,176.75)(15.500,2.000){2}{\rule{3.734pt}{0.600pt}}
\put(1024,179.75){\rule{7.709pt}{0.600pt}}
\multiput(1024.00,178.75)(16.000,2.000){2}{\rule{3.854pt}{0.600pt}}
\put(1056,182.25){\rule{6.350pt}{0.600pt}}
\multiput(1056.00,180.75)(17.820,3.000){2}{\rule{3.175pt}{0.600pt}}
\multiput(1087.00,185.99)(5.726,0.503){3}{\rule{4.950pt}{0.121pt}}
\multiput(1087.00,183.75)(21.726,4.000){2}{\rule{2.475pt}{0.600pt}}
\multiput(1119.00,189.99)(3.734,0.502){5}{\rule{3.870pt}{0.121pt}}
\multiput(1119.00,187.75)(22.968,5.000){2}{\rule{1.935pt}{0.600pt}}
\multiput(1150.00,194.99)(3.734,0.502){5}{\rule{3.870pt}{0.121pt}}
\multiput(1150.00,192.75)(22.968,5.000){2}{\rule{1.935pt}{0.600pt}}
\multiput(1181.00,199.99)(2.499,0.501){9}{\rule{2.893pt}{0.121pt}}
\multiput(1181.00,197.75)(25.996,7.000){2}{\rule{1.446pt}{0.600pt}}
\multiput(1213.00,206.99)(2.073,0.501){11}{\rule{2.475pt}{0.121pt}}
\multiput(1213.00,204.75)(25.863,8.000){2}{\rule{1.238pt}{0.600pt}}
\multiput(1244.00,214.99)(1.877,0.501){13}{\rule{2.283pt}{0.121pt}}
\multiput(1244.00,212.75)(27.261,9.000){2}{\rule{1.142pt}{0.600pt}}
\multiput(1276.00,224.00)(1.460,0.501){17}{\rule{1.841pt}{0.121pt}}
\multiput(1276.00,221.75)(27.179,11.000){2}{\rule{0.920pt}{0.600pt}}
\multiput(1307.00,235.00)(1.262,0.500){21}{\rule{1.627pt}{0.121pt}}
\multiput(1307.00,232.75)(28.623,13.000){2}{\rule{0.813pt}{0.600pt}}
\put(773.0,174.0){\rule[-0.300pt]{22.645pt}{0.600pt}}
\put(176,173){\usebox{\plotpoint}}
\put(679,172.25){\rule{7.468pt}{0.600pt}}
\multiput(679.00,171.75)(15.500,1.000){2}{\rule{3.734pt}{0.600pt}}
\put(176.0,173.0){\rule[-0.300pt]{121.173pt}{0.600pt}}
\put(773,173.25){\rule{7.468pt}{0.600pt}}
\multiput(773.00,172.75)(15.500,1.000){2}{\rule{3.734pt}{0.600pt}}
\put(710.0,174.0){\rule[-0.300pt]{15.177pt}{0.600pt}}
\put(836,174.25){\rule{7.468pt}{0.600pt}}
\multiput(836.00,173.75)(15.500,1.000){2}{\rule{3.734pt}{0.600pt}}
\put(867,175.75){\rule{7.709pt}{0.600pt}}
\multiput(867.00,174.75)(16.000,2.000){2}{\rule{3.854pt}{0.600pt}}
\put(899,178.25){\rule{6.350pt}{0.600pt}}
\multiput(899.00,176.75)(17.820,3.000){2}{\rule{3.175pt}{0.600pt}}
\put(930,181.25){\rule{6.550pt}{0.600pt}}
\multiput(930.00,179.75)(18.405,3.000){2}{\rule{3.275pt}{0.600pt}}
\multiput(962.00,184.99)(5.537,0.503){3}{\rule{4.800pt}{0.121pt}}
\multiput(962.00,182.75)(21.037,4.000){2}{\rule{2.400pt}{0.600pt}}
\multiput(993.00,188.99)(2.918,0.501){7}{\rule{3.250pt}{0.121pt}}
\multiput(993.00,186.75)(24.254,6.000){2}{\rule{1.625pt}{0.600pt}}
\multiput(1024.00,194.99)(2.499,0.501){9}{\rule{2.893pt}{0.121pt}}
\multiput(1024.00,192.75)(25.996,7.000){2}{\rule{1.446pt}{0.600pt}}
\multiput(1056.00,201.99)(1.817,0.501){13}{\rule{2.217pt}{0.121pt}}
\multiput(1056.00,199.75)(26.399,9.000){2}{\rule{1.108pt}{0.600pt}}
\multiput(1087.00,211.00)(1.508,0.501){17}{\rule{1.895pt}{0.121pt}}
\multiput(1087.00,208.75)(28.066,11.000){2}{\rule{0.948pt}{0.600pt}}
\multiput(1119.00,222.00)(1.130,0.500){23}{\rule{1.479pt}{0.121pt}}
\multiput(1119.00,219.75)(27.931,14.000){2}{\rule{0.739pt}{0.600pt}}
\multiput(1150.00,236.00)(0.983,0.500){27}{\rule{1.313pt}{0.121pt}}
\multiput(1150.00,233.75)(28.276,16.000){2}{\rule{0.656pt}{0.600pt}}
\multiput(1181.00,252.00)(0.849,0.500){33}{\rule{1.161pt}{0.121pt}}
\multiput(1181.00,249.75)(29.591,19.000){2}{\rule{0.580pt}{0.600pt}}
\multiput(1213.00,271.00)(0.676,0.500){41}{\rule{0.959pt}{0.121pt}}
\multiput(1213.00,268.75)(29.010,23.000){2}{\rule{0.479pt}{0.600pt}}
\multiput(1244.00,294.00)(0.641,0.500){45}{\rule{0.918pt}{0.120pt}}
\multiput(1244.00,291.75)(30.095,25.000){2}{\rule{0.459pt}{0.600pt}}
\multiput(1276.00,319.00)(0.552,0.500){51}{\rule{0.814pt}{0.120pt}}
\multiput(1276.00,316.75)(29.310,28.000){2}{\rule{0.407pt}{0.600pt}}
\multiput(1307.00,347.00)(0.532,0.500){55}{\rule{0.790pt}{0.120pt}}
\multiput(1307.00,344.75)(30.360,30.000){2}{\rule{0.395pt}{0.600pt}}
\put(804.0,175.0){\rule[-0.300pt]{7.709pt}{0.600pt}}
\end{picture}
\end{figure}
\newpage
\begin{figure}
% GNUPLOT: LaTeX picture
\setlength{\unitlength}{0.240900pt}
\ifx\plotpoint\undefined\newsavebox{\plotpoint}\fi
\sbox{\plotpoint}{\rule[-0.500pt]{1.000pt}{1.000pt}}%
\begin{picture}(1500,1350)(0,0)
\font\gnuplot=cmr10 at 10pt
\gnuplot
\sbox{\plotpoint}{\rule[-0.500pt]{1.000pt}{1.000pt}}%
\put(176.0,68.0){\rule[-0.500pt]{1.000pt}{303.293pt}}
\put(176.0,68.0){\rule[-0.500pt]{4.818pt}{1.000pt}}
\put(154,68){\makebox(0,0)[r]{$135$}}
\put(1416.0,68.0){\rule[-0.500pt]{4.818pt}{1.000pt}}
\put(176.0,316.0){\rule[-0.500pt]{4.818pt}{1.000pt}}
\put(154,316){\makebox(0,0)[r]{$140$}}
\put(1416.0,316.0){\rule[-0.500pt]{4.818pt}{1.000pt}}
\put(176.0,564.0){\rule[-0.500pt]{4.818pt}{1.000pt}}
\put(154,564){\makebox(0,0)[r]{$145$}}
\put(1416.0,564.0){\rule[-0.500pt]{4.818pt}{1.000pt}}
\put(176.0,812.0){\rule[-0.500pt]{4.818pt}{1.000pt}}
\put(154,812){\makebox(0,0)[r]{$150$}}
\put(1416.0,812.0){\rule[-0.500pt]{4.818pt}{1.000pt}}
\put(176.0,1059.0){\rule[-0.500pt]{4.818pt}{1.000pt}}
\put(154,1059){\makebox(0,0)[r]{$155$}}
\put(1416.0,1059.0){\rule[-0.500pt]{4.818pt}{1.000pt}}
\put(176.0,1307.0){\rule[-0.500pt]{4.818pt}{1.000pt}}
\put(154,1307){\makebox(0,0)[r]{$160$}}
\put(1416.0,1307.0){\rule[-0.500pt]{4.818pt}{1.000pt}}
\put(176.0,68.0){\rule[-0.500pt]{1.000pt}{4.818pt}}
\put(176,23){\makebox(0,0){$0$}}
\put(176.0,1307.0){\rule[-0.500pt]{1.000pt}{4.818pt}}
\put(455.0,68.0){\rule[-0.500pt]{1.000pt}{4.818pt}}
\put(455,23){\makebox(0,0){$50$}}
\put(455.0,1307.0){\rule[-0.500pt]{1.000pt}{4.818pt}}
\put(735.0,68.0){\rule[-0.500pt]{1.000pt}{4.818pt}}
\put(735,23){\makebox(0,0){$100$}}
\put(735.0,1307.0){\rule[-0.500pt]{1.000pt}{4.818pt}}
\put(1014.0,68.0){\rule[-0.500pt]{1.000pt}{4.818pt}}
\put(1014,23){\makebox(0,0){$150$}}
\put(1014.0,1307.0){\rule[-0.500pt]{1.000pt}{4.818pt}}
\put(1294.0,68.0){\rule[-0.500pt]{1.000pt}{4.818pt}}
\put(1294,23){\makebox(0,0){$200$}}
\put(1294.0,1307.0){\rule[-0.500pt]{1.000pt}{4.818pt}}
\put(176.0,68.0){\rule[-0.500pt]{303.534pt}{1.000pt}}
\put(1436.0,68.0){\rule[-0.500pt]{1.000pt}{303.293pt}}
\put(176.0,1327.0){\rule[-0.500pt]{303.534pt}{1.000pt}}
\put(847,-30){\makebox(0,0)[r]{$T(MeV)$}}
\put(1950,-179){\makebox(0,0)[r]{{\bf{Figure 3: Tempearture dependence of $\pi$-meson mass; the nomenclature is same as fig.1.}}}}
\put(64,688){\makebox(0,0)[r]{$m_\pi^*(MeV)$}}
\put(1204,1059){\makebox(0,0)[r]{$W$}}
\put(1422,395){\makebox(0,0)[r]{$ZM$}}
\put(1422,455){\makebox(0,0)[r]{$ZM2$}}
\put(1422,554){\makebox(0,0)[r]{$ZM3$}}
\put(176.0,68.0){\rule[-0.500pt]{1.000pt}{303.293pt}}
\sbox{\plotpoint}{\rule[-0.300pt]{0.600pt}{0.600pt}}%
\put(176,316){\usebox{\plotpoint}}
\multiput(847.00,316.99)(3.232,0.502){5}{\rule{3.390pt}{0.121pt}}
\multiput(847.00,314.75)(19.964,5.000){2}{\rule{1.695pt}{0.600pt}}
\put(176.0,316.0){\rule[-0.300pt]{161.644pt}{0.600pt}}
\multiput(930.00,321.99)(3.357,0.502){5}{\rule{3.510pt}{0.121pt}}
\multiput(930.00,319.75)(20.715,5.000){2}{\rule{1.755pt}{0.600pt}}
\put(874.0,321.0){\rule[-0.300pt]{13.490pt}{0.600pt}}
\multiput(986.00,326.99)(3.357,0.502){5}{\rule{3.510pt}{0.121pt}}
\multiput(986.00,324.75)(20.715,5.000){2}{\rule{1.755pt}{0.600pt}}
\multiput(1014.00,331.99)(3.357,0.502){5}{\rule{3.510pt}{0.121pt}}
\multiput(1014.00,329.75)(20.715,5.000){2}{\rule{1.755pt}{0.600pt}}
\multiput(1042.00,337.00)(1.457,0.501){15}{\rule{1.830pt}{0.121pt}}
\multiput(1042.00,334.75)(24.202,10.000){2}{\rule{0.915pt}{0.600pt}}
\multiput(1070.00,346.99)(3.357,0.502){5}{\rule{3.510pt}{0.121pt}}
\multiput(1070.00,344.75)(20.715,5.000){2}{\rule{1.755pt}{0.600pt}}
\multiput(1098.00,352.00)(1.018,0.500){23}{\rule{1.350pt}{0.121pt}}
\multiput(1098.00,349.75)(25.198,14.000){2}{\rule{0.675pt}{0.600pt}}
\multiput(1126.00,366.00)(0.703,0.500){35}{\rule{0.990pt}{0.121pt}}
\multiput(1126.00,363.75)(25.945,20.000){2}{\rule{0.495pt}{0.600pt}}
\multiput(1155.00,385.00)(0.500,0.534){51}{\rule{0.120pt}{0.793pt}}
\multiput(1152.75,385.00)(28.000,28.354){2}{\rule{0.600pt}{0.396pt}}
\multiput(1183.00,415.00)(0.500,1.063){51}{\rule{0.120pt}{1.414pt}}
\multiput(1180.75,415.00)(28.000,56.065){2}{\rule{0.600pt}{0.707pt}}
\multiput(1211.00,474.00)(0.500,11.285){51}{\rule{0.120pt}{13.436pt}}
\multiput(1208.75,474.00)(28.000,592.114){2}{\rule{0.600pt}{6.718pt}}
\multiput(1239.00,1094.00)(0.500,1.518){51}{\rule{0.120pt}{1.950pt}}
\multiput(1236.75,1094.00)(28.000,79.953){2}{\rule{0.600pt}{0.975pt}}
\multiput(1266.99,1178.00)(0.502,1.099){5}{\rule{0.121pt}{1.350pt}}
\multiput(1264.75,1178.00)(5.000,7.198){2}{\rule{0.600pt}{0.675pt}}
\multiput(1271.99,1188.00)(0.501,0.871){7}{\rule{0.121pt}{1.150pt}}
\multiput(1269.75,1188.00)(6.000,7.613){2}{\rule{0.600pt}{0.575pt}}
\multiput(1277.00,1198.99)(0.472,0.502){5}{\rule{0.750pt}{0.121pt}}
\multiput(1277.00,1196.75)(3.443,5.000){2}{\rule{0.375pt}{0.600pt}}
\put(958.0,326.0){\rule[-0.300pt]{6.745pt}{0.600pt}}
\put(176,316){\usebox{\plotpoint}}
\multiput(847.00,316.99)(3.232,0.502){5}{\rule{3.390pt}{0.121pt}}
\multiput(847.00,314.75)(19.964,5.000){2}{\rule{1.695pt}{0.600pt}}
\put(176.0,316.0){\rule[-0.300pt]{161.644pt}{0.600pt}}
\multiput(930.00,321.99)(3.357,0.502){5}{\rule{3.510pt}{0.121pt}}
\multiput(930.00,319.75)(20.715,5.000){2}{\rule{1.755pt}{0.600pt}}
\put(874.0,321.0){\rule[-0.300pt]{13.490pt}{0.600pt}}
\multiput(986.00,326.99)(3.357,0.502){5}{\rule{3.510pt}{0.121pt}}
\multiput(986.00,324.75)(20.715,5.000){2}{\rule{1.755pt}{0.600pt}}
\multiput(1014.00,331.99)(3.357,0.502){5}{\rule{3.510pt}{0.121pt}}
\multiput(1014.00,329.75)(20.715,5.000){2}{\rule{1.755pt}{0.600pt}}
\multiput(1042.00,336.99)(3.357,0.502){5}{\rule{3.510pt}{0.121pt}}
\multiput(1042.00,334.75)(20.715,5.000){2}{\rule{1.755pt}{0.600pt}}
\multiput(1070.00,341.99)(3.357,0.502){5}{\rule{3.510pt}{0.121pt}}
\multiput(1070.00,339.75)(20.715,5.000){2}{\rule{1.755pt}{0.600pt}}
\multiput(1098.00,346.99)(3.357,0.502){5}{\rule{3.510pt}{0.121pt}}
\multiput(1098.00,344.75)(20.715,5.000){2}{\rule{1.755pt}{0.600pt}}
\multiput(1126.00,351.99)(1.636,0.501){13}{\rule{2.017pt}{0.121pt}}
\multiput(1126.00,349.75)(23.814,9.000){2}{\rule{1.008pt}{0.600pt}}
\multiput(1154.00,361.00)(1.457,0.501){15}{\rule{1.830pt}{0.121pt}}
\multiput(1154.00,358.75)(24.202,10.000){2}{\rule{0.915pt}{0.600pt}}
\multiput(1182.00,371.00)(1.457,0.501){15}{\rule{1.830pt}{0.121pt}}
\multiput(1182.00,368.75)(24.202,10.000){2}{\rule{0.915pt}{0.600pt}}
\multiput(1210.00,381.00)(1.457,0.501){15}{\rule{1.830pt}{0.121pt}}
\multiput(1210.00,378.75)(24.202,10.000){2}{\rule{0.915pt}{0.600pt}}
\multiput(1238.00,391.00)(1.457,0.501){15}{\rule{1.830pt}{0.121pt}}
\multiput(1238.00,388.75)(24.202,10.000){2}{\rule{0.915pt}{0.600pt}}
\multiput(1266.00,401.00)(0.947,0.500){25}{\rule{1.270pt}{0.121pt}}
\multiput(1266.00,398.75)(25.364,15.000){2}{\rule{0.635pt}{0.600pt}}
\put(958.0,326.0){\rule[-0.300pt]{6.745pt}{0.600pt}}
\put(176,316){\usebox{\plotpoint}}
\multiput(847.00,316.99)(3.232,0.502){5}{\rule{3.390pt}{0.121pt}}
\multiput(847.00,314.75)(19.964,5.000){2}{\rule{1.695pt}{0.600pt}}
\put(176.0,316.0){\rule[-0.300pt]{161.644pt}{0.600pt}}
\multiput(930.00,321.99)(3.357,0.502){5}{\rule{3.510pt}{0.121pt}}
\multiput(930.00,319.75)(20.715,5.000){2}{\rule{1.755pt}{0.600pt}}
\put(874.0,321.0){\rule[-0.300pt]{13.490pt}{0.600pt}}
\multiput(986.00,326.99)(3.357,0.502){5}{\rule{3.510pt}{0.121pt}}
\multiput(986.00,324.75)(20.715,5.000){2}{\rule{1.755pt}{0.600pt}}
\multiput(1014.00,331.99)(3.357,0.502){5}{\rule{3.510pt}{0.121pt}}
\multiput(1014.00,329.75)(20.715,5.000){2}{\rule{1.755pt}{0.600pt}}
\multiput(1042.00,336.99)(3.357,0.502){5}{\rule{3.510pt}{0.121pt}}
\multiput(1042.00,334.75)(20.715,5.000){2}{\rule{1.755pt}{0.600pt}}
\multiput(1070.00,341.99)(3.357,0.502){5}{\rule{3.510pt}{0.121pt}}
\multiput(1070.00,339.75)(20.715,5.000){2}{\rule{1.755pt}{0.600pt}}
\multiput(1098.00,346.99)(1.636,0.501){13}{\rule{2.017pt}{0.121pt}}
\multiput(1098.00,344.75)(23.814,9.000){2}{\rule{1.008pt}{0.600pt}}
\multiput(1126.00,355.99)(3.357,0.502){5}{\rule{3.510pt}{0.121pt}}
\multiput(1126.00,353.75)(20.715,5.000){2}{\rule{1.755pt}{0.600pt}}
\multiput(1154.00,361.00)(1.457,0.501){15}{\rule{1.830pt}{0.121pt}}
\multiput(1154.00,358.75)(24.202,10.000){2}{\rule{0.915pt}{0.600pt}}
\multiput(1182.00,371.00)(0.947,0.500){25}{\rule{1.270pt}{0.121pt}}
\multiput(1182.00,368.75)(25.364,15.000){2}{\rule{0.635pt}{0.600pt}}
\multiput(1210.00,386.00)(1.457,0.501){15}{\rule{1.830pt}{0.121pt}}
\multiput(1210.00,383.75)(24.202,10.000){2}{\rule{0.915pt}{0.600pt}}
\multiput(1238.00,396.00)(0.947,0.500){25}{\rule{1.270pt}{0.121pt}}
\multiput(1238.00,393.75)(25.364,15.000){2}{\rule{0.635pt}{0.600pt}}
\multiput(1267.00,410.00)(0.500,0.807){51}{\rule{0.120pt}{1.114pt}}
\multiput(1264.75,410.00)(28.000,42.687){2}{\rule{0.600pt}{0.557pt}}
\put(958.0,326.0){\rule[-0.300pt]{6.745pt}{0.600pt}}
\put(176,316){\usebox{\plotpoint}}
\multiput(847.00,316.99)(3.232,0.502){5}{\rule{3.390pt}{0.121pt}}
\multiput(847.00,314.75)(19.964,5.000){2}{\rule{1.695pt}{0.600pt}}
\put(176.0,316.0){\rule[-0.300pt]{161.644pt}{0.600pt}}
\multiput(930.00,321.99)(3.357,0.502){5}{\rule{3.510pt}{0.121pt}}
\multiput(930.00,319.75)(20.715,5.000){2}{\rule{1.755pt}{0.600pt}}
\put(874.0,321.0){\rule[-0.300pt]{13.490pt}{0.600pt}}
\multiput(986.00,326.99)(3.357,0.502){5}{\rule{3.510pt}{0.121pt}}
\multiput(986.00,324.75)(20.715,5.000){2}{\rule{1.755pt}{0.600pt}}
\multiput(1014.00,331.99)(3.357,0.502){5}{\rule{3.510pt}{0.121pt}}
\multiput(1014.00,329.75)(20.715,5.000){2}{\rule{1.755pt}{0.600pt}}
\multiput(1042.00,336.99)(3.357,0.502){5}{\rule{3.510pt}{0.121pt}}
\multiput(1042.00,334.75)(20.715,5.000){2}{\rule{1.755pt}{0.600pt}}
\multiput(1070.00,342.00)(1.457,0.501){15}{\rule{1.830pt}{0.121pt}}
\multiput(1070.00,339.75)(24.202,10.000){2}{\rule{0.915pt}{0.600pt}}
\multiput(1098.00,351.99)(4.971,0.503){3}{\rule{4.350pt}{0.121pt}}
\multiput(1098.00,349.75)(18.971,4.000){2}{\rule{2.175pt}{0.600pt}}
\multiput(1126.00,356.00)(1.457,0.501){15}{\rule{1.830pt}{0.121pt}}
\multiput(1126.00,353.75)(24.202,10.000){2}{\rule{0.915pt}{0.600pt}}
\multiput(1154.00,366.00)(0.947,0.500){25}{\rule{1.270pt}{0.121pt}}
\multiput(1154.00,363.75)(25.364,15.000){2}{\rule{0.635pt}{0.600pt}}
\multiput(1182.00,381.00)(1.121,0.500){25}{\rule{1.470pt}{0.121pt}}
\multiput(1182.00,378.75)(29.949,15.000){2}{\rule{0.735pt}{0.600pt}}
\multiput(1215.00,396.00)(0.773,0.500){25}{\rule{1.070pt}{0.121pt}}
\multiput(1215.00,393.75)(20.779,15.000){2}{\rule{0.535pt}{0.600pt}}
\multiput(1239.00,410.00)(0.500,0.807){51}{\rule{0.120pt}{1.114pt}}
\multiput(1236.75,410.00)(28.000,42.687){2}{\rule{0.600pt}{0.557pt}}
\multiput(1267.00,455.00)(0.500,1.609){51}{\rule{0.120pt}{2.057pt}}
\multiput(1264.75,455.00)(28.000,84.730){2}{\rule{0.600pt}{1.029pt}}
\put(958.0,326.0){\rule[-0.300pt]{6.745pt}{0.600pt}}
\end{picture}
\end{figure}
\end{document}